\documentclass[letter]{spie}

\usepackage[utf8]{inputenc}
\usepackage{graphicx}
\usepackage{siunitx}
\usepackage{capt-of}
\usepackage{array}
\usepackage{caption}
\usepackage{subcaption}
\usepackage{amsmath,amssymb}
\usepackage{booktabs}       % professional-quality tables
\usepackage{makecell}
\usepackage{arydshln}       % using dashed lines
\usepackage{caption}
\usepackage[export]{adjustbox}

\usepackage{cleveref}

\title{Unsupervised Super-Resolution: Creating High-Resolution Medical Images from Low-Resolution Anisotropic Examples}

\author{%
	Jörg Sander\supit{a}\supit{b}, Bob D. de Vos\supit{a}\supit{b} and Ivana Išgum\supit{a}\supit{b}\supit{c}%
	\skiplinehalf%
	\small%
	\supit{a}Department of Biomedical Engineering and Physics, Amsterdam University Medical Centers, University of Amsterdam, The Netherlands\\%
	\supit{b}Amsterdam Cardiovascular Sciences, Amsterdam University Medical Centers, University of Amsterdam, The Netherlands\\%
	\supit{c}Department of Radiology and Nuclear Medicine, Amsterdam University Medical Centers, University of Amsterdam, The Netherlands\\%
}

\authorinfo{Send correspondence to J.Sander (email: jsander@amsterdamumc.nl)}

\begin{document}
	\maketitle 
	
	\begin{abstract}
		Although high resolution isotropic \num{3}D medical images are desired in clinical practice, their acquisition is not always feasible. Instead, lower resolution images are upsampled to higher resolution using conventional interpolation methods. Sophisticated learning-based super-resolution approaches are frequently unavailable in clinical setting, because such methods require training with high-resolution isotropic examples. To address this issue, we propose a learning-based super-resolution approach that can be trained using solely anisotropic images, i.e. without high-resolution ground truth data. The method exploits the latent space, generated by autoencoders trained on anisotropic images, to increase spatial resolution in low-resolution images. The method was trained and evaluated using \num{100} publicly available cardiac cine MR scans from the Automated Cardiac Diagnosis Challenge (ACDC). The quantitative results show that the proposed method performs better than conventional interpolation methods. Furthermore, the qualitative results indicate that especially finer cardiac structures are synthesized with high quality. The method has the potential to be applied to other anatomies and modalities and can be easily applied to any \num{3}D anisotropic medical image dataset. 
	\end{abstract}    
	
	
	\section{Introduction}
	
	High spatial resolution of medical images is considered a very important quality component for accurate disease diagnosis and prognosis. However, acquiring high-resolution images is not always feasible. For example, MR acquisition is coupled with a trade-off between signal-to-noise ratio, spatial resolution, temporal resolution and acquisition time. As a result, time-series, like cardiac cine magnetic resonance images (CMRI), are typically highly anisotropic with in-plane resolution in the millimeter range and through-plane resolution in the range of tens of millimeters. When isotropic images are required for (semi-)automatic analysis, the acquired data are typically upsampled using conventional interpolation methods like Linear, B-spline, and Lanczos resampling.
	
	Although these methods are generally considered fast and broadly applicable, they only leverage intensity information from one image dimension to increase spatial or temporal resolution. 
	In contrast, learning-based super-resolution (SR) approaches\cite{glasner2009super, kim2010single, yang2013fast, ledig2017photo, timofte2018ntire, bhatia2014super, oktay2016multi, oktay2017anatomically, chen2018brain, basty2018super, pham2019multiscale} exploit all image dimensions to recover the high spatial frequency information that was lost in the image acquisition stage. Image super-resolution is an active research topic in the computer vision community \cite{glasner2009super, kim2010single, yang2013fast, ledig2017photo, timofte2018ntire} as well as in the field of medical image processing \cite{bhatia2014super, oktay2016multi, oktay2017anatomically, chen2018brain, basty2018super, pham2019multiscale}. However, to train such methods, high-resolution training examples are required, which are in clinical practice hardly at hand. 
	
	\begin{figure}
		\captionsetup[subfigure]{justification=centering}
		\centering
		\includegraphics[width=3.1in]{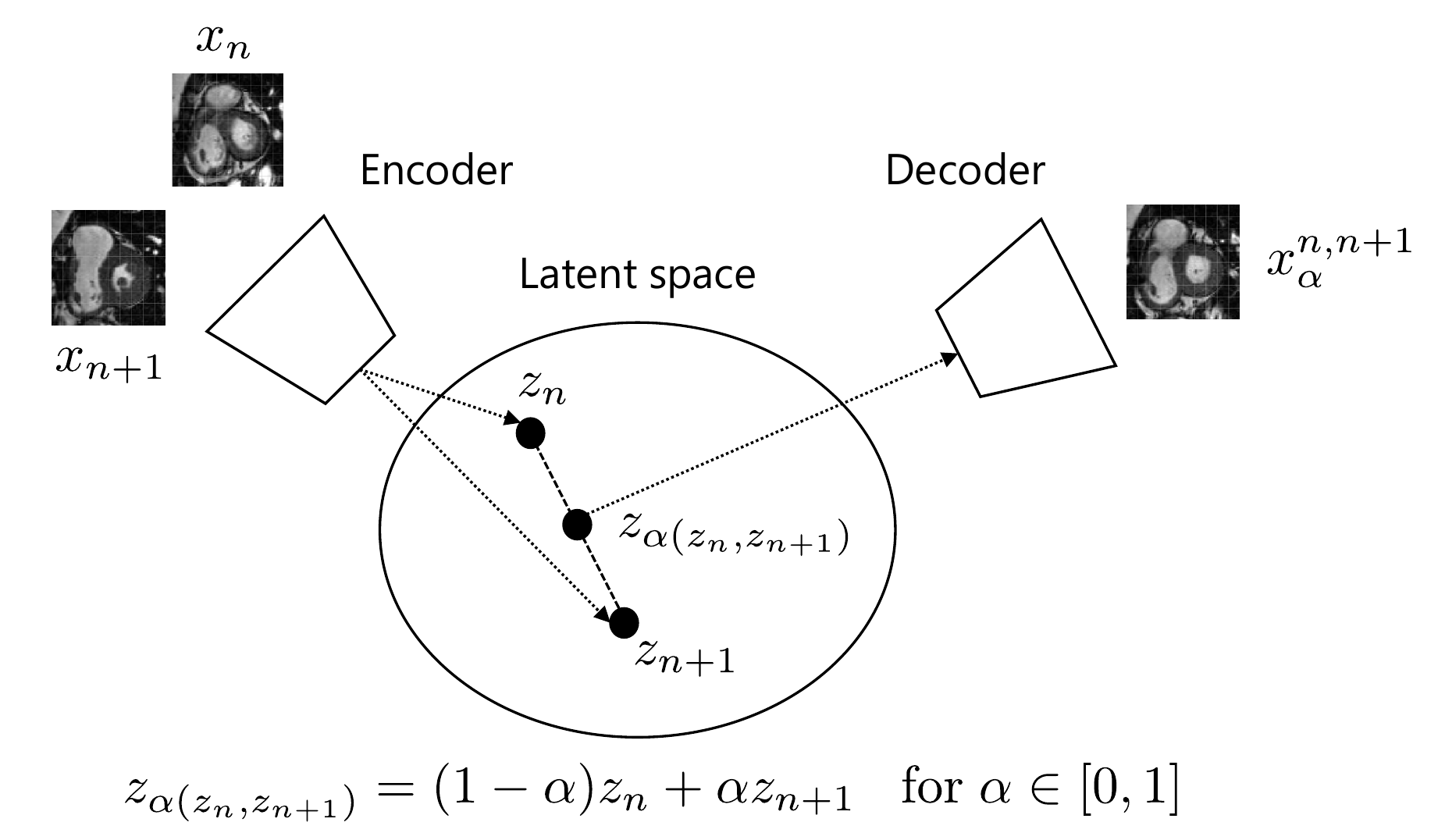}		
		\vspace{2ex}
		\caption{Visualization of proposed super-resolution method. We exploit the latent space interpolation ability of autoencoders to perform super-resolution in anisotropic medical images.}	
		\label{fig_method}
	\end{figure}
	
	Several methods \cite{weigert2017isotropic, jog2016self, zhao2018self} have been proposed to increase resolution of images without using high-resolution training data. Weigert et al. \cite{weigert2017isotropic} proposed a super-resolution deep-learning approach applicable to anisotropic fluorescence microscopy images only. Jog et al. \cite{jog2016self} used a Fourier-based method that combines multiple augmentations of a single input image to estimate Fourier coefficients of higher frequency ranges absent in the anisotropic images. Zhao et al. \cite{zhao2018self} used the previous approach to combine two upsampled images that were created with a self-supervised super-resolution approach. This approach learns a low-to-high-resolution mapping from synthesized anisotropic versions of isotropic in-plane images. Although the latter two approaches might be applicable for highly-anisotropic images, they were evaluated on brain MRI with relatively mild anisotropy (\num{2}$-\SI{3}{\milli\meter}$ through-plane).
	
	% 	The method consists of multiple stages and is complicated to apply. Inspired by the work of \cite{jog2016self}, Zhao et al. \cite{zhao2018self} proposed a self-supervised super-resolution approach that exploits high resolution in-plane images and generates a low-resolution version to find the low-resolution to high-resolution mapping. However, the latter two approaches were evaluated on brain MRI with relative mild anisotropy (\num{2}$-\SI{3}{\milli\meter}$ through-plane).
	
	We propose an unsupervised learning-based super-resolution approach and evaluate it on highly anisotropic cardiac MR images. The method is able to create high-resolution \num{3}D images from low-resolution images. Specifically, we upsample highly anisotropic images by using the latent space interpolation abilities of an autoencoder. The method encodes high-resolution in-plane information and uses the encodings to increase resolution in through-plane direction. Unlike previous super-resolution methods using deep learning, our method can be trained in an end-to-end fashion using merely the interpolation ability of an autoencoder trained on anisotropic images. Moreover, our approach can be straightforwardly applied to any imaging modality. We compare our method to conventional interpolation methods and show its superior performance.
	
	\section{Data description}
	
	Cardiac cine MR images from the MICCAI \num{2017} Automated Cardiac Diagnosis Challenge (ACDC) \cite{bernard2018deep} were used. The dataset consists of short-axis CMRIs from 100 patients uniformly distributed over normal cardiac function and four disease groups: dilated cardiomyopathy, hypertrophic cardiomyopathy, heart failure with infarction, and right ventricular abnormality. Detailed acquisition protocol is described by Bernard et al.~\cite{bernard2018deep}. %Briefly, short-axis CMRIs were acquired with two MRI scanners of different magnetic strengths (\num{1.5} and \num{3.0} T). Images were made during breath hold using a conventional steady-state free precession (SSFP) sequence. 
	Briefly, CMRIs have an in-plane resolution ranging from \num{1.37} to \SI{1.68}{\milli\meter} (average reconstruction matrix \num{243} $\times$ \num{217} voxels) with slice spacing varying from \num{5} to \SI{10}{\milli\meter}. Per patient 28 to 40 time points are provided covering the cardiac cycle. Each volume consists of on average ten slices covering the heart. To correct for intensity differences among scans, image intensities of each volume were rescaled and clamped between [\num{0}, \num{1}] based on their \num{1}$^{st}$ and \num{99}$^{th}$ percentiles. Furthermore, to correct for differences in-plane voxel sizes, image slices were resampled to \num{1.4}$\times\SI{1.4}{\milli\meter}^2$.
	
	\section{Method}
	The proposed method trains an autoencoder to compress and reconstruct high-resolution \num{2}D slices taken from an anisotropic \num{3}D cardiac MRI dataset. The trained autoencoder is used to generate new slices in unseen images to increase spatial resolution. New intermediate slices are generated mixing the latent space encodings of two spatially adjacent slices. Subsequently, the mixed encoding is decoded to the intermediate slice. Figure~\ref{fig_method} visualizes the approach. %  resolution of the same volumes can be achieved by mixing latent representation codes of two spatially adjacent slices and decoding the result.  
	
	An autoencoder is an unsupervised learning algorithm that aims to learn a lower dimensional representation of the input. It consists of an encoder and decoder implemented as neural network. The encoder compresses the input into a lower dimensional space, also referred to as latent space representation, which captures the most salient features of the input. The decoder uses the latent space representation to generate an approximate reconstruction of the input. In general, training an autoencoder aims to minimize the dissimilarity between the input and the corresponding reconstruction.
	
	To generate super-resolution images, in this work, the trained autoencoder is used to project two spatially adjacent slices ($x^n$, $x^{n+1}$) onto a lower dimensional latent space. It is assumed that through-plane resolution is low and in-plane resolution is high. Thereafter, latent representations $z_n$ and $z_{n+1}$ are combined using a convex combination $z_{\alpha(z_n, z_{n+1})} = (1 - \alpha) z_n + \alpha z_{n+1}$ for $\alpha \in [0, 1]$. Finally, using the decoder a new slice $x_{\alpha}^{n, n+1}$ is generated by decoding the mixture of latent codes. Increasing $\alpha$ from \num{0} to \num{1} results in a sequence of new slices where each subsequent slice is progressively less semantically similar to $x^n$ and more semantically similar to $x^{n+1}$. The obtained stack of slices can therefore be used to increase the spatial resolution in the direction from which the slices were extracted. 
	
	\begin{figure}
		\captionsetup[subfigure]{justification=centering}
		\centering
		\subfloat[Mid-cavity cardiac MR slices ]{\includegraphics[width=2.5in]{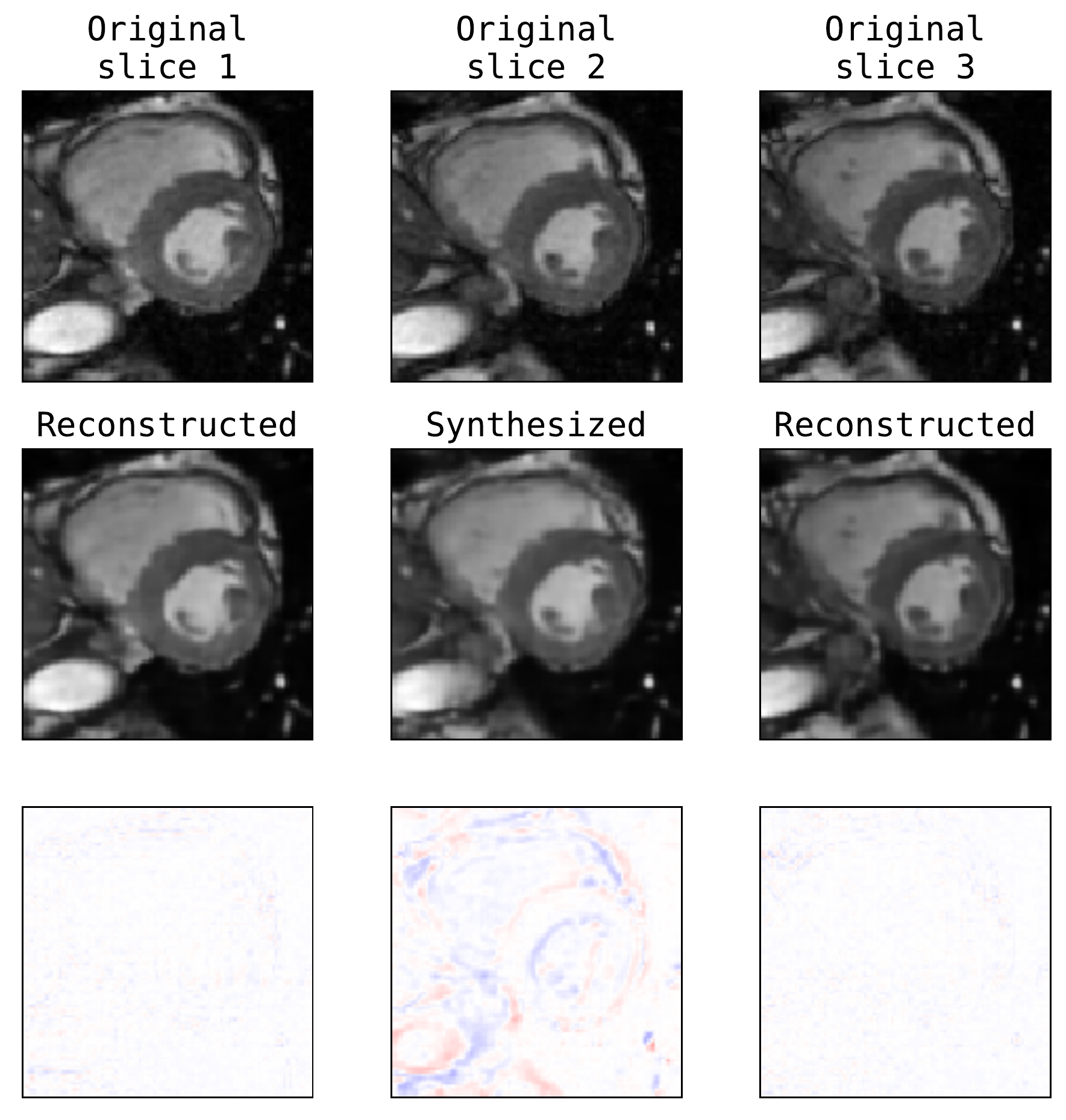}  
			\label{fig_qualitative_example_1a}  }  \hspace{12ex}
		\vspace{1ex}
		\subfloat[ Mid-cavity cardiac MR slices]{\includegraphics[width=2.5in]{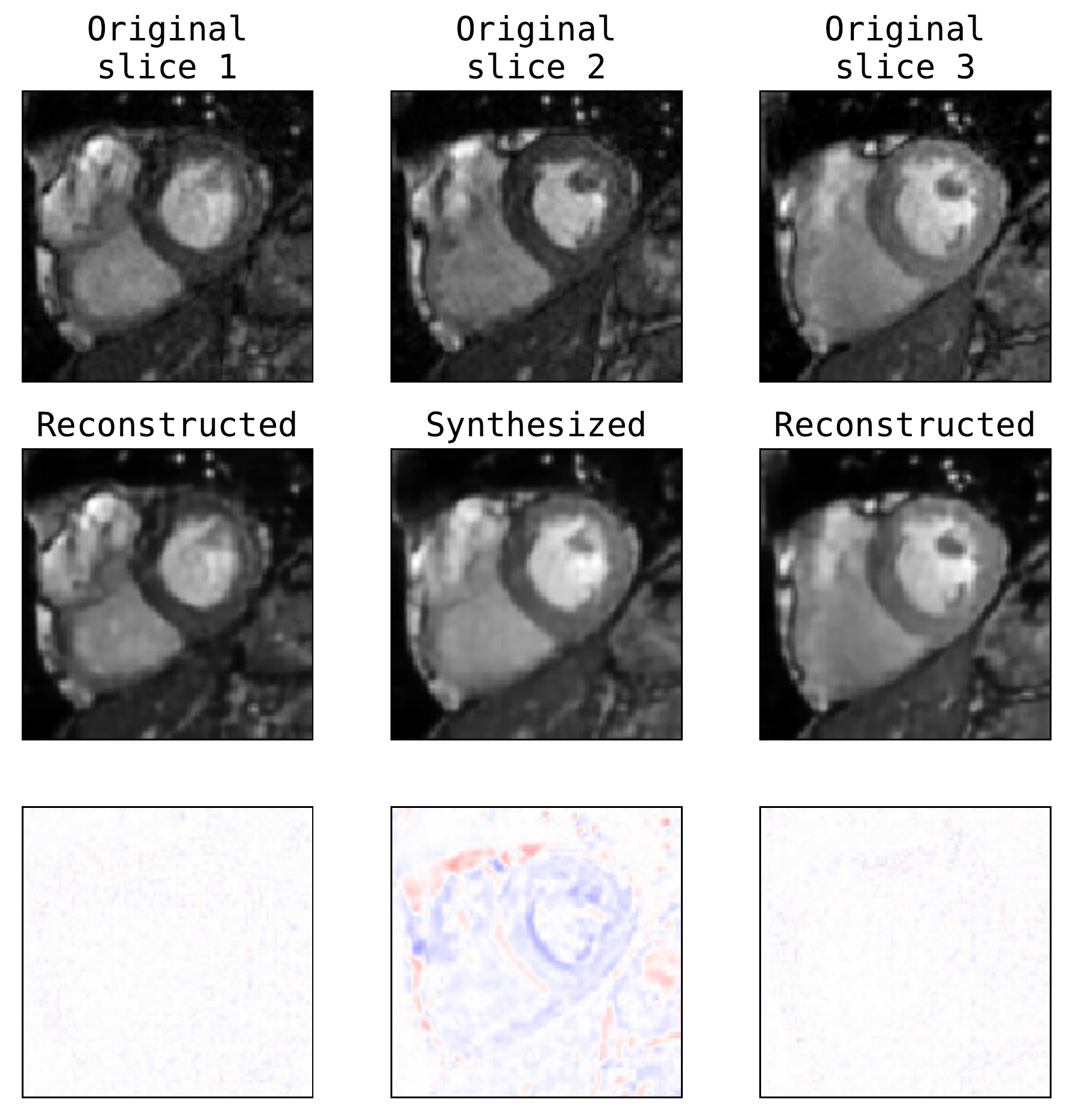}  
			\label{fig_qualitative_example_1b}  }  % \vspace{2ex} }  
		\vspace{1ex}
		
		\subfloat[Apical cardiac MR slices]{\includegraphics[width=2.5in]{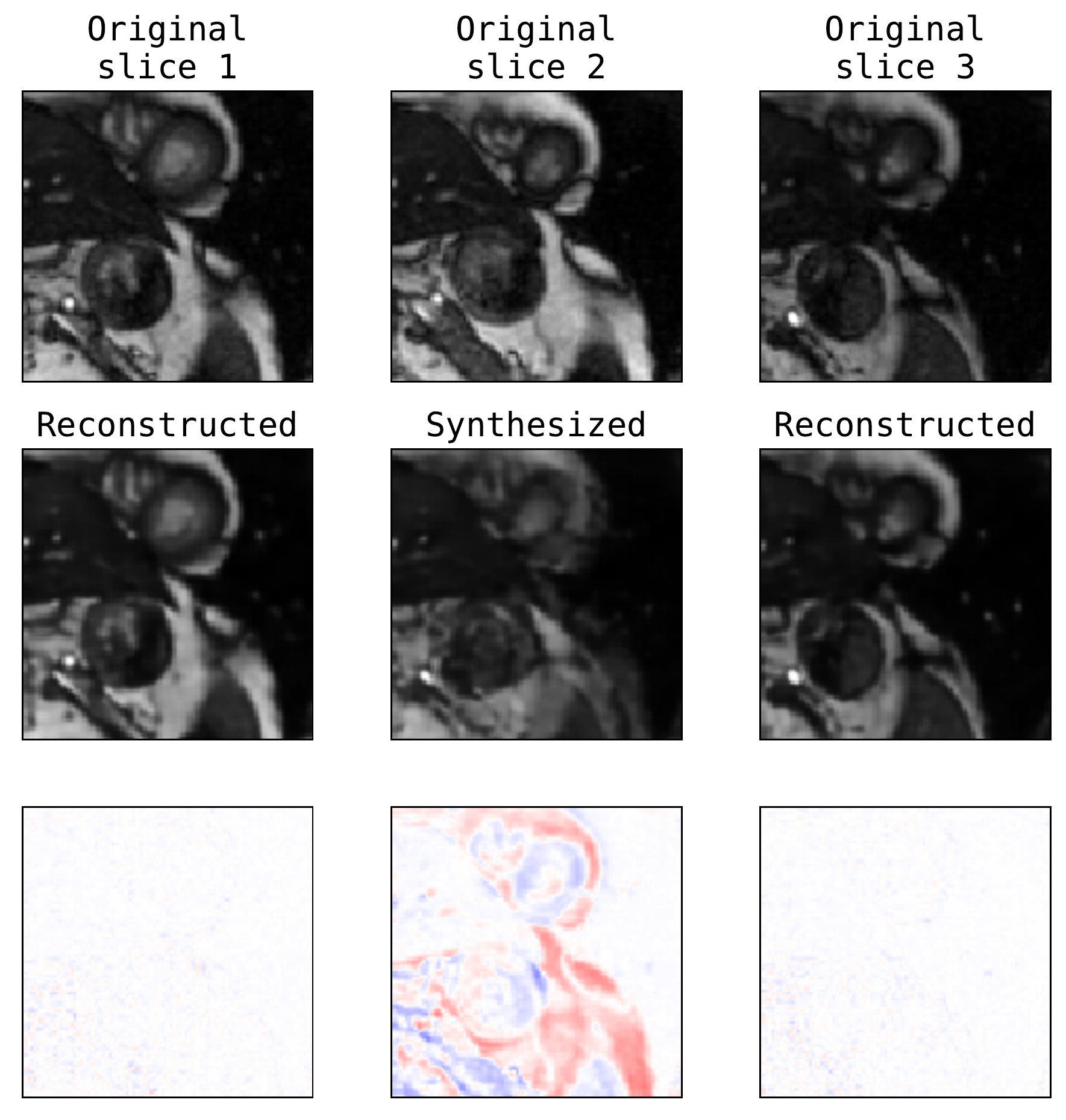}  
			\label{fig_qualitative_example_1c} }  \hspace{12ex} 
		\vspace{1ex}
		\subfloat[Basal cardiac MR slices]{\includegraphics[width=2.5in]{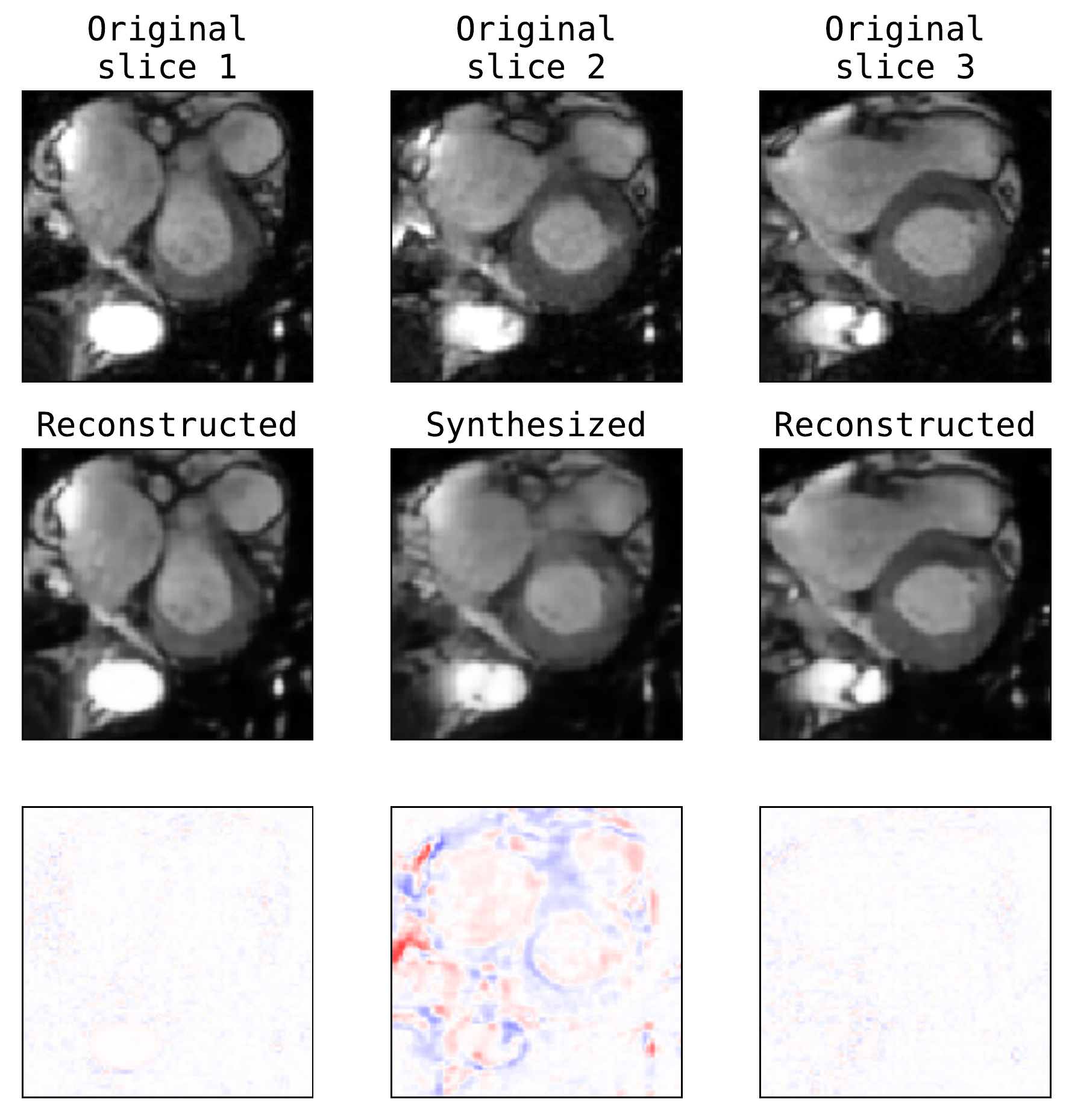}  
			\label{fig_qualitative_example_1d}  }
		\vspace{1ex} 
		
		\caption{Four examples of (a) and (b) Mid-Cavity and (c) Apical (d) Basal cardiac cine MR image slices. For each of the sub-figures the following applies. Top: Three adjacent input image slices. Middle: Reconstructions of the input by the autoencoder (left and right), and synthesized image by the autoencoder in between two input slices (middle). Bottom: Differences between input images and corresponding reconstructed or synthesized slices (blue corresponds to negative and red to positive differences).}
		
		\label{fig_qualitative_example}
	\end{figure}
	
	\section{Experiments and results}
	Experiments generating CMR images with super-resolution were performed using the ACDC dataset. The lowest resolution anisotropic images (\num{7.2}$-\SI{10}{\milli\meter}$ through-plane) were split into training (70) and validation (18) sets. The \num{12} remaining higher resolution anisotropic images ($\SI{5}{\milli\meter}$ through-plane resolution) were used as test set for quantitative as well as qualitative evaluation. 
	
	The architecture of the encoder consists of four blocks, each with two consecutive \num{3}$\times$\num{3} convolutional layers, followed by batch normalization and \num{2}$\times$\num{2} average pooling. The first block uses \num{32} kernels and the number of kernels is doubled in the first convolutional layer of all subsequent blocks. The last block is followed by two additional \num{3}$\times$\num{3} convolutional layers of \num{256}, and \num{16} kernels for the final output layer. The output of the final convolutional layer is used as latent space representation of the input. All convolutional layers except for the final use a leaky ReLU nonlinearity. The architecture of the decoder is reverse of the encoder. It consists of four blocks of two consecutive \num{3}$\times$\num{3} convolutional layers with leaky ReLU	nonlinearities followed by batch normalization and \num{2}$\times$\num{2} nearest neighbor upsampling. The number of kernels is halved after each upsampling layer. The last block is followed by two additional \num{3}$\times$\num{3} convolutional layers. To preserve input size all convolutions (encoder and decoder) use zero-padding of size \num{1}. The model was implemented using the PyTorch framework.
	
	To train the autoencoder patches of \num{128}$\times$\num{128} pixels were randomly chosen from the training set and augmented by \num{90} degree rotations of the images and random intensity changes. The mean squared error was used as a loss and weights were optimized using the Adam optimizer with learning rate of \num{1e-5}. Test images were center-cropped to \num{128}$\times$\num{128} pixels covering all cardiac structures of interest. To evaluate our method, we mimicked $\SI{10}{\milli\meter}$ through-plane resolution by excluding every other slice in the test images. These excluded slices were subsequently recovered by synthesizing them using our approach. %All experiments were implemented using PyTorch \cite{paszke2017automatic}.
	
	\begin{figure}
		\captionsetup[subfigure]{justification=centering}
		\centering
		\includegraphics[width=.65\textwidth]{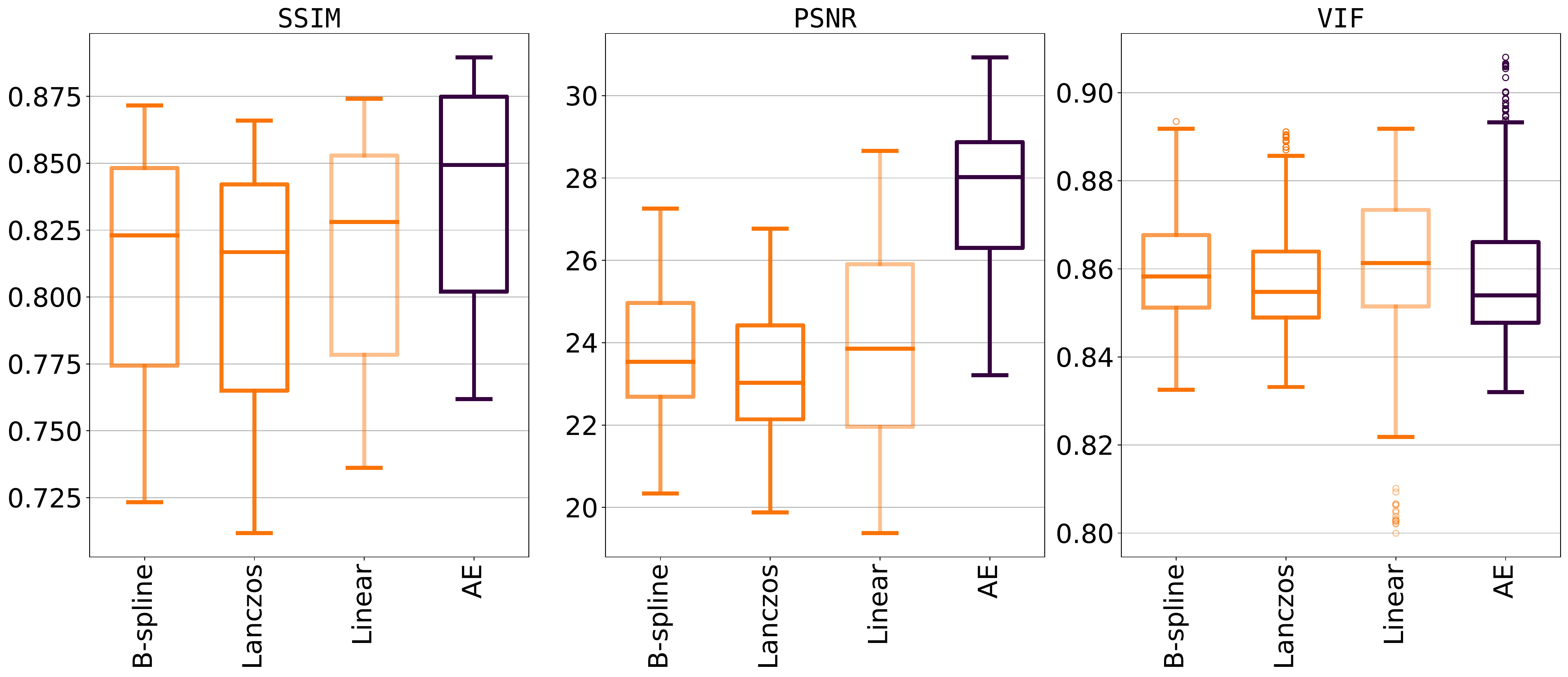}
		
		\caption{Boxplots showing evaluation of super-resolution performance on cardiac MRI for conventional interpolation methods (B-Spline, Lanczos, Linear) compared to proposed method (AE) in terms of SSIM, PSNR and Visual Information Fidelity (VIF). A higher score indicates better performance. }	
		\label{fig_boxplot_cardiac}
	\end{figure}
	
	\subsection{Qualitative evaluation}
	Figure~\ref{fig_qualitative_example} visualizes qualitative results for the proposed method. The results show that the trained autoencoder is able to reconstruct high-quality images i.e. input slices. This is best visible in the trabeculae of the left ventricle. Furthermore, visual evaluation reveals that synthesized slices, i.e. those that are generated by the autoencoder to increase the spatial resolution, show an anatomically and semantically meaningful transition between the two neighboring slices. Moreover, one can observe that slices synthesized at the mid-cavity anatomy contain fewer errors than synthesized slices at the apex or base of the heart. Furthermore, deviations from the reference image seem to be more pronounced for the right ventricle compared to the left ventricle. % as acquired and reconstructed at the scanner
	
	\begin{figure}[ht!]
		\captionsetup[subfigure]{justification=centering}
		\centering
		
		\subfloat[]{\includegraphics[width=.56\textwidth]{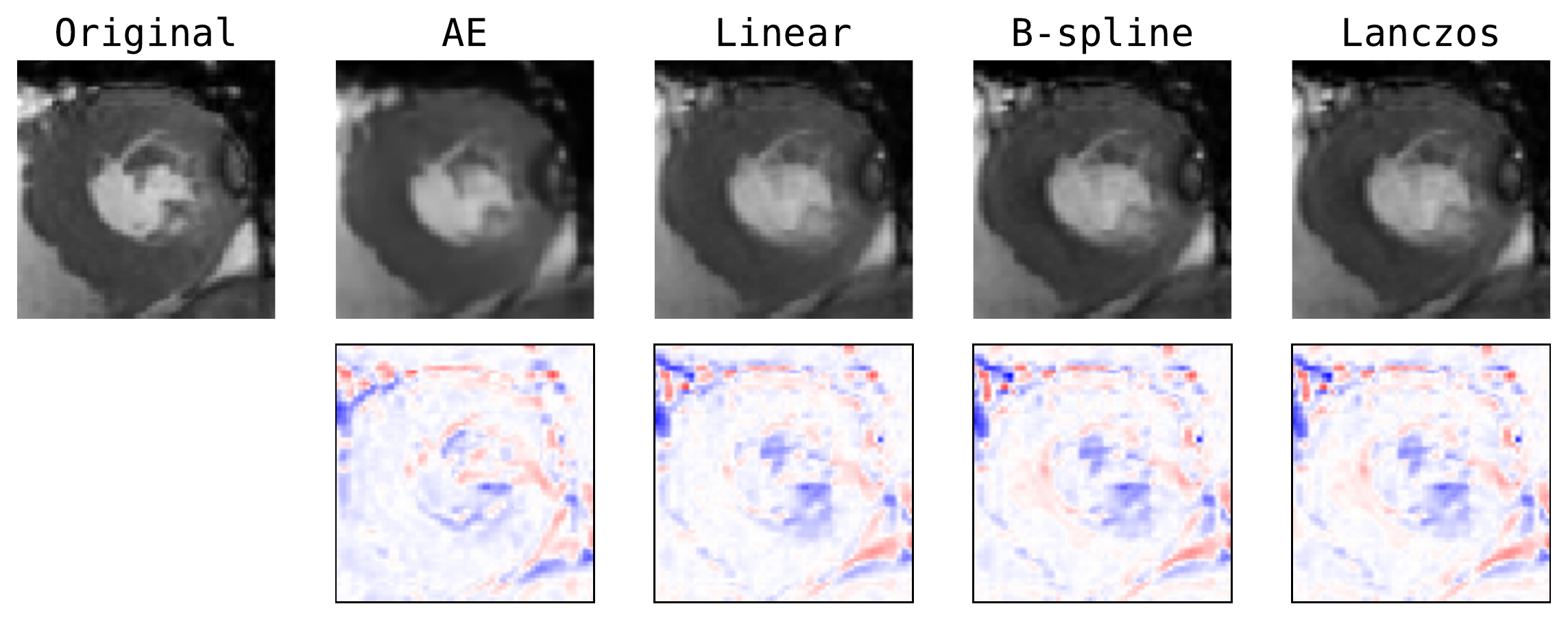} 
			%\vspace{1ex}
			%\hspace{4ex}
			\label{fig_qual_comparison_ex1}
		}
		
		\subfloat[]{\includegraphics[width=.56\textwidth]{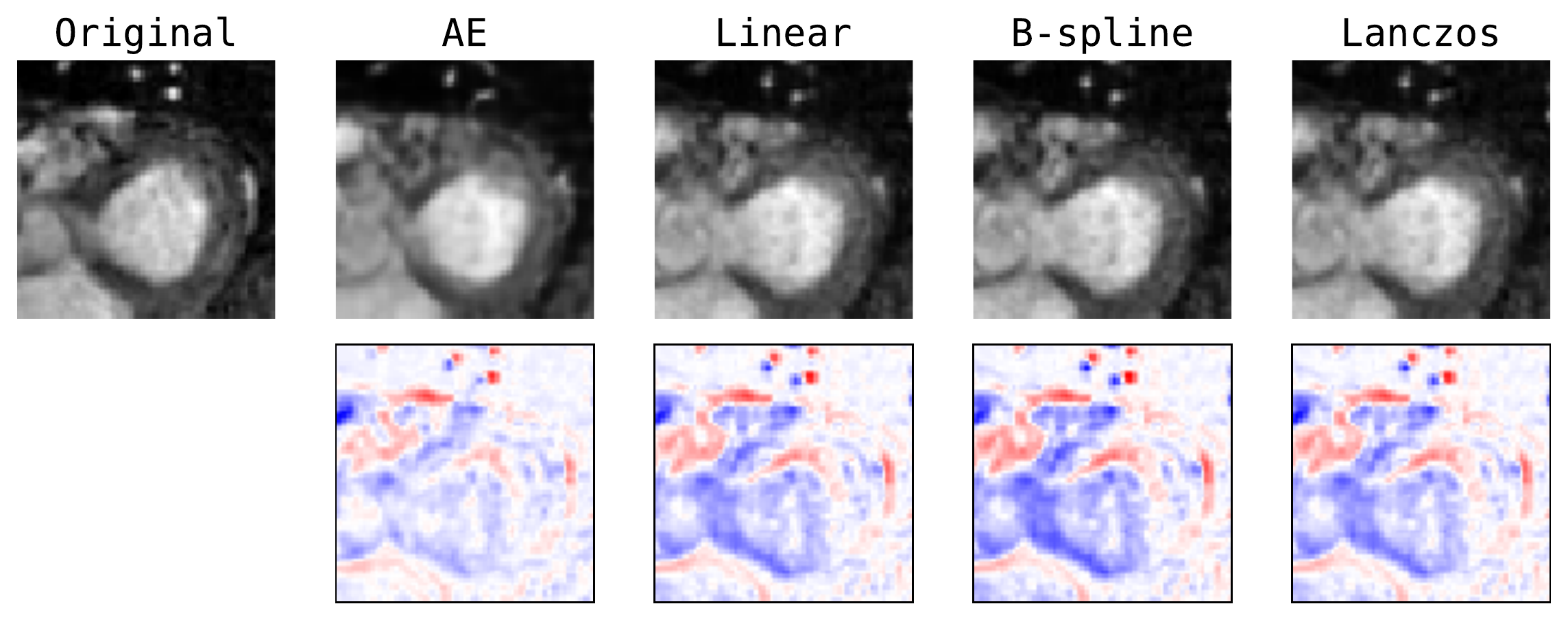} 
			%\vspace{1ex}
			\label{fig_qual_comparison_ex2}
		}
		
		\subfloat[]{\includegraphics[width=.56\textwidth]{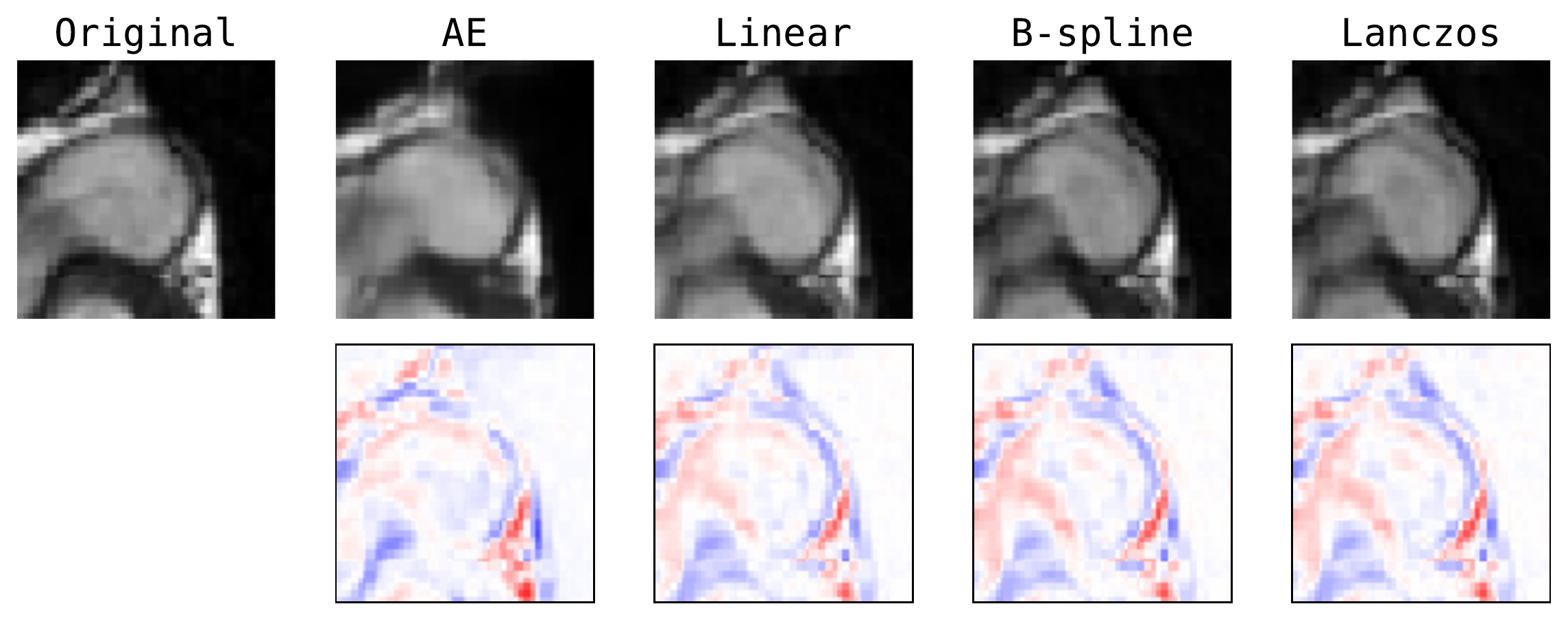} %\hspace{4ex}  
			\label{fig_qual_comparison_ex3} }
		
		\subfloat[]{\includegraphics[width=.56\textwidth]{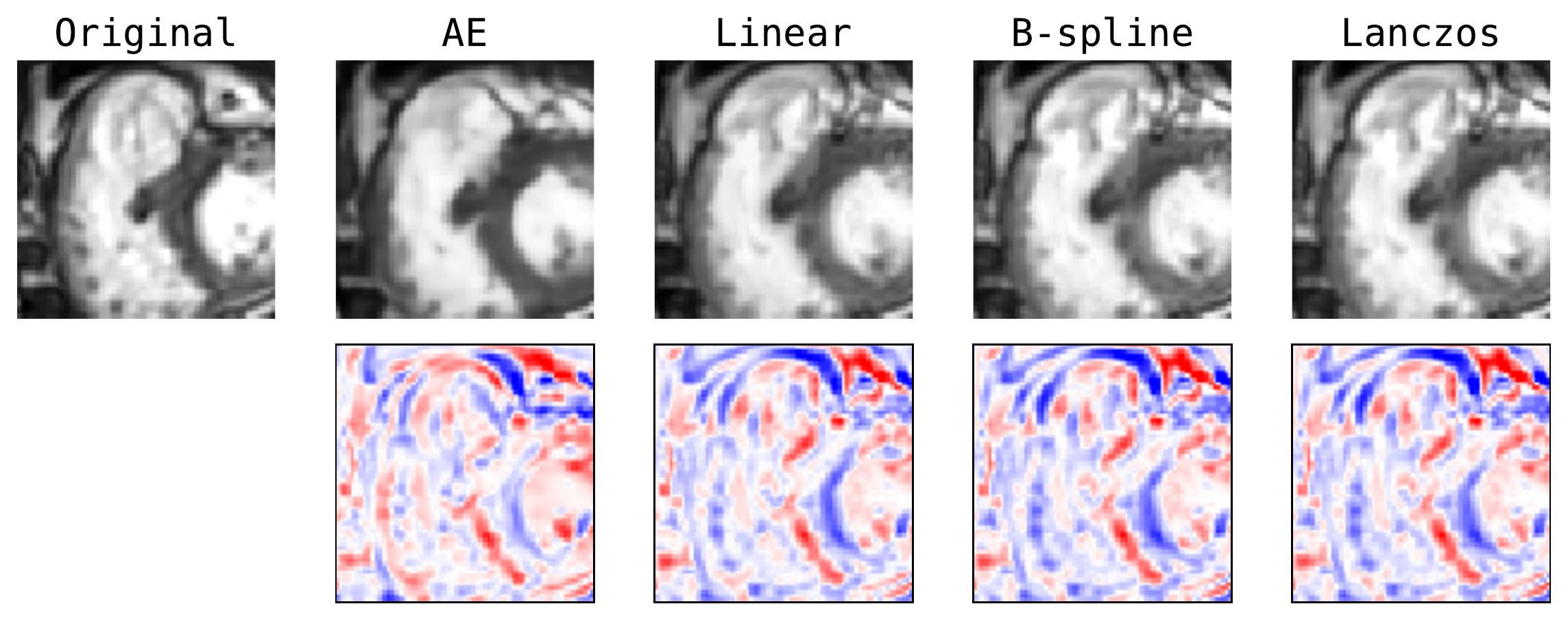} 
			
			\label{fig_qual_comparison_ex4} }
		
		\caption{Four examples of (a) and (b) left ventricle and (c) and (d) right ventricle cine MR image comparing slices generated with our approach and conventional interpolation methods. The following applies to each sub-figure. Top row shows reference image (first column) and images generated by different conventional interpolation methods (second to last column). Bottom row visualizes difference between synthesized/interpolated and reference images (blue corresponds to negative and red to positive differences).}
		\label{fig_compare_with_conventional}	
	\end{figure}
	
	\subsection{Comparison with conventional interpolation methods}
	Additionally, super-resolution performance was quantitatively evaluated and compared with Linear, B-spline and Lanczos resampling in terms of Structural Similarity Index Measure (SSIM), Peak Signal-to-Noise Ratio (PSNR) and Visual Information Fidelity (VIF)\cite{sheikh2005information}. The results are shown in Figure~\ref{fig_boxplot_cardiac}. We observe that the proposed super-resolution method achieves better performance when evaluated by SSIM and PSNR compared with the conventional interpolation methods. These differences are statistically significant ($p \lll 0.001$) using the one-sided Wilcoxon signed-rank test. Performance differences in terms of VIF are negligible. Qualitative comparison of the methods shown in Figure~\ref{fig_compare_with_conventional} reveals that synthesized images generated by the proposed super-resolution method contain fewer errors, especially for the left ventricle myocardium and its fine trabeculae structures.
	
	\section{Discussion and conclusions}
	We have proposed a learning-based super-resolution approach that is based on autoencoders. The method is able to create high-resolution \num{3}D cardiac MR images using only anisotropic images for training. Hence, our approach does not require high-resolution isotropic ground truth data volumes. The performance of the method is affected by large variation of anatomy between adjacent slices. This might be mitigated by increasing training data set size. Nevertheless, the results indicate that our method can exploit the latent space of autoencoders to generate smooth and semantically meaningful interpolations in image space. Moreover, the results on cardiac MR images reveal that our approach outperforms conventional interpolation methods. 
	
	Related super-resolution methods \cite{jog2016self, zhao2018self} that do not require high-resolution training data were evaluated on brain MRI with relatively mild anisotropy. In contrast, our approach was evaluated on highly anisotropic cardiac MRI. Nonetheless, evaluating our approach on a similar brain MRI set will be addressed in our future work. Additionally, since our method learns from the data itself, it can be readily applied to other modalities. 
	
	To conclude, in this work we have proposed an unsupervised learning-based super-resolution approach that can create high-resolution cardiac MR images from highly anisotropic CMRIs.
	
	\section{NEW OR BREAKTHROUGH WORK TO BE PRESENTED}
	We have presented an unsupervised learning-based super-resolution approach to increase spatial resolution in anisotropic medical images that does not rely on high-resolution ground truth data. 
	
	\vspace{2ex}
	\noindent This work has not been submitted elsewhere.
	
	\acknowledgments
	\noindent This study was performed within the DLMedIA program (P15-26) funded by Dutch Technology Foundation with participation of PIE Medical Imaging.
	
	\bibliography{sr_bibliography}
	\bibliographystyle{spiebib}
	
\end{document}